\documentclass[a4paper,11pt,oneside]{article}

\usepackage[english]{babel}
\usepackage[T1]{fontenc}
\usepackage[utf8]{inputenc}

\usepackage[pdftex,unicode]{hyperref}
\hypersetup{pdftitle=gasmodel: An R Package for Generalized Autoregressive Score Models}
\hypersetup{pdfauthor=Vladimír Holý}

\usepackage{xcolor}
\definecolor{mycolor}{rgb}{0.50,0.00,0.00}
\hypersetup{colorlinks=true, linkcolor=mycolor, anchorcolor=mycolor, citecolor=mycolor, filecolor=mycolor, urlcolor=mycolor}

\usepackage[margin=60pt]{geometry}
\usepackage{amsmath}
\usepackage{amssymb}
\usepackage{amsthm}

\usepackage[group-minimum-digits=3]{siunitx}

\usepackage{graphicx}
\usepackage{float}
\usepackage{hhline}
\usepackage{multirow}
\usepackage{booktabs}
\usepackage{dcolumn}

\usepackage[authoryear]{natbib}

\usepackage{verbatim}

\newcommand{\pkg}[1]{\textbf{#1}}
\makeatletter
\newcommand\code{\bgroup\@makeother\_\@makeother\~\@makeother\$\@codex}
\def\@codex#1{{\normalfont\ttfamily\hyphenchar\font=-1 #1}\egroup}
\makeatother
\newenvironment{longcode}[0]{\footnotesize\verbatim}{\endverbatim}

\begin{document}

\begin{center}
{\Large \bfseries gasmodel: An R Package for Generalized Autoregressive \\ Score Models}
\end{center}

\begin{center}
{\bfseries Vladimír Holý} \\
Prague University of Economics and Business \\
Winston Churchill Square 1938/4, 130 67 Prague 3, Czechia \\
\href{mailto:vladimir.holy@vse.cz}{vladimir.holy@vse.cz} \\
\end{center}

\noindent
\textbf{Abstract:}
Generalized autoregressive score (GAS) models are a class of observation-driven time series models that employ the score to dynamically update time-varying parameters of the underlying probability distribution. GAS models have been extensively studied and numerous variants have been proposed in the literature to accommodate diverse data types and probability distributions. This paper introduces the gasmodel package, which has been designed to facilitate the estimation, forecasting, and simulation of a wide range of GAS models. The package provides a rich selection of distributions, offers flexible options for specifying dynamics, and allows to incorporate exogenous variables. Model estimation utilizes the maximum likelihood method.
\\

\\

\\

\section{Introduction}
\label{sec:intro}

%% About GAS
The generalized autoregressive score (GAS) models, introduced by \citet{Creal2013} and \citet{Harvey2013}, have emerged as a valuable and contemporary framework for time series modeling. These models, also referred to as dynamic conditional score (DCS) models or score-driven models, offer flexibility by accommodating various underlying probability distributions and time-varying parameters. GAS models are observation-driven, effectively capturing the dynamic behavior of time-varying parameters through the autoregressive term and the score, i.e., the gradient of the log-likelihood function. The GAS framework enables the formulation of a wide range of dynamic models suitable for many commonly used univariate and multivariate data types.

% Other R GAS Packages
There are several packages and code available in R that handle GAS models. One notable package is \pkg{GAS} developed by \citet{Ardia2018} and \citet{Ardia2019}, which provides functionality for both univariate and multivariate GAS models. The current version of the \pkg{GAS} package, 0.3.4, supports 16 distributions. However, the model specification in the \pkg{GAS} package is somewhat limited, only allowing for basic dynamics without the inclusion of exogenous variables. Additionally, this package lacks distributions for certain more specialized data types, such as circular, compositional, and ranking data. The package thus supports only a limited selection of GAS models found in the literature. For a more detailed comparison of the \pkg{gasmodel} and \pkg{GAS} packages, see the \code{comparison_gas} vignette from the \pkg{gasmodel} package. Another relevant R package is \pkg{betategarch} by \cite{Sucarrat2013}, which deals specifically with the Beta-Skew-t-EGARCH model, a GAS model for time-varying volatility based on the Student's t-distribution. In Python, the \pkg{PyFlux} library by \cite{Taylor2017} deals with time series analysis and features various GAS models including the Beta-Skew-t-EGARCH model, standard GAS models, GAS random walk models, GAS pairwise comparison models, and GAS regression models. In Julia, the \pkg{ScoreDrivenModels.jl} package by \cite{Bodin2020} provides a framework for standard GAS models. The \pkg{Time Series Lab} program by \cite{Lit2021} is a stand-alone GUI application designed to model and forecast time series, including standard GAS models, GAS pairwise comparison models, and GAS regression models. Additional R, Python, MATLAB, and Ox code for some specific GAS models, often associated with individual research papers, can be found on the \url{www.gasmodel.com} website.

% Main Features of the Package
In this paper, we present the \pkg{gasmodel} package, which is designed to provide comprehensive functionality that encompasses a wide range of GAS models documented in the existing literature. It offers versatile model specification and core features available for the entire spectrum of implemented distributions. The current version of the \pkg{gasmodel} package, 0.6.2, offers a selection of 36 distributions, catering to various univariate and multivariate data types such as binary, categorical, ranking, count, integer, circular, interval, compositional, duration, and real data types. A comprehensive list of these distributions is provided in Table \ref{tab:distr}. For details on each distribution, see the \code{distributions} vignette from the \pkg{gasmodel} package. Model specification within the package allows for flexible customization, enabling users to incorporate different parametrizations, exogenous variables, joint and separate modeling of exogenous variables and dynamics, higher score and autoregressive orders, custom and unconditional initial values of time-varying parameters, fixed and bounded values of coefficients, and missing values. Model estimation is performed by the maximum likelihood method. Standard errors of coefficients are estimated using the empirical Hessian matrix. Furthermore, the package offers a range of functionalities including forecasting, simulation, bootstrapping, and assessment of parameter uncertainty. Comprehensive documentation is provided with the package, offering details on each distribution and its corresponding parametrizations.

% Where to Find It
The \pkg{gasmodel} package is accessible on CRAN at \url{cran.r-project.org/package=gasmodel}. Additionally, users can find the development version of the package on GitHub at \url{github.com/vladimirholy/gasmodel}, providing them with the opportunity to report any bugs or issues they encounter.

% Paper Structure
The rest of the paper is as follows. In Section \ref{sec:gas}, we outline the key characteristics of GAS models. In Section \ref{sec:pkg}, we present an overview of the \pkg{gasmodel} package. In Section \ref{sec:case}, we present a case study demonstrating the practical application of the package. We conclude the paper in Section \ref{sec:con}.

\begin{table}
\begin{center}
\caption{List of available distributions and their parametrizations. First parametrization is the default.}
\label{tab:distr}
\small
\begin{tabular}{lllll}
\toprule
Label & Distribution & Dimension & Data Type & Parametrizations \\
\midrule
\code{alaplace} & Asymmetric Laplace & Uni. & Real & \code{meanscale}\\
\code{bernoulli} & Bernoulli & Uni. & Binary & \code{prob}\\
\code{beta} & Beta & Uni. & Interval & \code{conc, meansize, meanvar}\\
\code{bisa} & Birnbaum-Saunders & Uni. & Duration & \code{scale}\\
\code{burr} & Burr & Uni. & Duration & \code{scale}\\
\addlinespace
\code{cat} & Categorical & Multi. & Categorical & \code{worth}\\
\code{dirichlet} & Dirichlet & Multi. & Compositional & \code{conc}\\
\code{dpois} & Double Poisson & Uni. & Count & \code{mean}\\
\code{explog} & Exponential-Logarithmic & Uni. & Duration & \code{rate}\\
\code{exp} & Exponential & Uni. & Duration & \code{scale, rate}\\
\addlinespace
\code{fisk} & Fisk & Uni. & Duration & \code{scale}\\
\code{gamma} & Gamma & Uni. & Duration & \code{scale, rate}\\
\code{ged} & Generalized Error & Uni. & Real & \code{meanscale}\\
\code{gengamma} & Generalized Gamma & Uni. & Duration & \code{scale, rate}\\
\code{geom} & Geometric & Uni. & Count & \code{mean, prob}\\
\addlinespace
\code{kuma} & Kumaraswamy & Uni. & Interval & \code{conc}\\
\code{laplace} & Laplace & Uni. & Real & \code{meanscale}\\
\code{logistic} & Logistic & Uni. & Real & \code{meanscale}\\
\code{logitnorm} & Logit-Normal & Uni. & Interval & \code{logitmeanvar}\\
\code{lognorm} & Log-Normal & Uni. & Duration & \code{logmeanvar}\\
\addlinespace
\code{lomax} & Lomax & Uni. & Duration & \code{scale}\\
\code{mvnorm} & Multivariate Normal & Multi. & Real & \code{meanvar}\\
\code{mvt} & Multivariate Student’s t & Multi. & Real & \code{meanvar}\\
\code{negbin} & Negative Binomial & Uni. & Count & \code{nb2, prob}\\
\code{norm} & Normal & Uni. & Real & \code{meanvar}\\
\addlinespace
\code{pluce} & Plackett-Luce & Multi. & Ranking & \code{worth}\\
\code{pois} & Poisson & Uni. & Count & \code{mean}\\
\code{rayleigh} & Rayleigh & Uni. & Duration & \code{scale}\\
\code{skellam} & Skellam & Uni. & Integer & \code{meanvar, diff, meandisp}\\
\code{t} & Student’s t & Uni. & Real & \code{meanvar}\\
\addlinespace
\code{vonmises} & von Mises & Uni. & Circular & \code{meanconc}\\
\code{weibull} & Weibull & Uni. & Duration & \code{scale, rate}\\
\code{zigeom} & Zero-Inflated Geometric & Uni. & Count & \code{mean}\\
\code{zinegbin} & Zero-Inflated Negative Binomial & Uni. & Count & \code{nb2}\\
\code{zipois} & Zero-Inflated Poisson & Uni. & Count & \code{mean}\\
\addlinespace
\code{ziskellam} & Zero-Inflated Skellam & Uni. & Integer & \code{meanvar, diff, meandisp}\\
\bottomrule

\end{tabular}
\end{center}
\end{table}

\section{Generalized Autoregressive Score Models}
\label{sec:gas}

\subsection{Background}

The concept of utilizing the score as a driving mechanism for dynamics in time series was independently developed at both Vrije Universiteit Amsterdam and the University of Cambridge. At Vrije Universiteit Amsterdam, researchers established a comprehensive general methodology that encompasses various models driven by the score, known as the generalized autoregressive score (GAS) models. The initial findings were presented in the working paper \cite{Creal2008}, which was subsequently published as \cite{Creal2013}. At the University of Cambridge, the initial focus was on a specific model that employed the Student's t-distribution with dynamic volatility, named Beta-t-(E)GARCH. This approach was introduced in a working paper \cite{Harvey2008}. The book by \cite{Harvey2013} explores a variety of dynamic location and scale models driven by the score, referring to them as dynamic conditional score (DCS) models. Both \cite{Creal2013} and \cite{Harvey2013} are widely recognized as seminal contributions to the literature on GAS models. More recently, in order to reconcile different terminologies used in the literature, the term ``score-driven models'' has also emerged as a synonymous label.

The Scopus database reports 799 articles containing phrase ``generalized autoregressive score'' or ``dynamic conditional score'', as of August 17, 2025. The website \url{www.gasmodel.com} lists 438 articles, working papers, and books on GAS models, as of August 11, 2025.

\subsection{Basic Notation}

The goal is to model time series $y_t$, $t=1,\ldots,T$, which can be univariate or multivariate, continuous or discrete. Let $f_t$ denote the vector of time-varying parameters and $g$ the vector of static parameters. Let $p(y_t|f_t,g)$ denote the density function in the case of a continuous variable, or the probability mass function in the case of a discrete variable.

Constructing a model involves two main components: selecting an appropriate distribution and specifying the dynamics of its time-varying parameters.

\subsection{Score as the Key Ingredient}

In GAS models, the key ingredient driving the dynamics of the parameter vector $f_t$ is the score, i.e., the gradient of the log-likelihood function,
\begin{equation}
\nabla(y_t, f_t) = \frac{\partial \ln p(y_t | f_t, g)}{\partial f_t}.
\end{equation}
The score has zero expected value and its variance is known as the Fisher information,
\begin{equation}
\mathcal{I}(f_t) = \mathrm{E} \left[ \left( \frac{\partial \ln p(y_t | f_t, g)}{\partial f_t} \right)^2 \middle| f_t, g \right].
\end{equation}

The score quantifies the discrepancy between the fitted distribution, determined by $f_t$, and a particular observation $y_t$. As such, it can be employed as a correction term following the realization of an observation. When the score is positive, it suggests that the parameter of interest should be increased to better accommodate the observed data. Conversely, when the score is negative, decreasing the parameter would help in aligning the distribution with the observation. When the score is zero, it indicates that the current parameter value represents the optimal fit for the specific observation at hand.

An advantage of the score is that it takes into account the shape of the distribution. To illustrate this point, \cite{Creal2013} consider two GARCH models: one based on the normal distribution and another based on the Student's t-distribution. Now, imagine an extreme observation occurs. Due to its heavier tails, the Student's t-distribution assigns a higher probability to such extreme observations compared to the normal distribution. Crucially, this distinction is also mirrored in the score. Specifically, when assuming the normal distribution, the score for the extreme observation will have a significantly higher absolute value compared to when assuming the Student's t-distribution. The dynamics can thus reflect the shape of the distribution.

The simple difference between expectation and realization, commonly used as a correction term in various time series models, may not always be effective for distributions with specific support. \cite{Harvey2024} highlight this limitation in the context of circular time series. To illustrate this, let us suppose the expected value of an angle is 0.01 radians, but the actual observation turns out to be 6.27 radians. Although the numerical difference between these values is substantial, their corresponding angles are very similar as 0 radians and $2 \pi$ radians represent the exact same angle. This discrepancy highlights the inadequacy of using a simple difference metric. On the other hand, the score respects the circular nature of the data. For instance, when working with the von Mises distribution characterized by a time-varying location parameter $\mu_t$ and a static concentration parameter $\nu$, the score for $\mu_t$ is equal to $\nu\sin(y_t - \mu_t)$. By employing the sine function, the score accounts for the circularity of the data and ensures that the angular differences are appropriately considered during the analysis.

\subsection{Dynamics of Time Varying Parameters}

In GAS models, time-varying parameters $f_{t}$ follow the recursion
\begin{equation}
\label{eq:dynBasic}
f_{t} = \omega + \sum_{j=1}^P \alpha_j S(f_{t - j}) \nabla(y_{t - j}, f_{t - j}) + \sum_{k=1}^Q \varphi_k f_{t-k},
\end{equation}
where $\omega$ is the intercept, $\alpha_j$ are the score parameters, $\varphi_k$ are the autoregressive parameters, and $S(f_t)$ is a scaling function for the score. In the case of a single time-varying parameter, all these quantities are scalar. In the case of multiple time-varying parameters, $\omega$ and $\nabla(y_{t - j}, f_{t - j})$ are vectors, $\alpha_j$ and $\varphi_k$ are diagonal matrices, and $S(f_t)$ is a square matrix. In the majority of empirical studies, it is common practice to set the score order $P$ and the autoregressive order $Q$ to 1. Furthermore, one of three scaling functions is typically chosen: the unit function, the inverse of the Fisher information, or the square root of the inverse of the Fisher information. When the latter is used, the scaled score has unit variance. However, the choice of the scaling function is not always a straightforward task and is closely tied to the underlying distribution.

The dynamics of the model can be expanded to incorporate exogenous variables as
\begin{equation}
\label{eq:dynJoint}
f_{t} = \omega + \sum_{i=1}^M \beta_i x_{ti} + \sum_{j=1}^P \alpha_j S(f_{t - j}) \nabla(y_{t - j}, f_{t - j}) + \sum_{k=1}^Q \varphi_k f_{t-k},
\end{equation}
where $\beta_i$ are the regression parameters associated with the exogenous variables $x_{ti}$. Alternatively, a different model can be obtained by defining the recursion in the fashion of regression models with dynamic errors as
\begin{equation}
\label{eq:dynSep}
f_{t} = \omega + \sum_{i=1}^M \beta_i x_{ti} + e_{t}, \quad e_t = \sum_{j=1}^P \alpha_j S(f_{t - j}) \nabla(y_{t - j}, f_{t - j}) + \sum_{k=1}^Q \varphi_k e_{t-k}.
\end{equation}
The key distinction between the two models lies in the impact of exogenous variables on $f_t$. Specifically, in the former model formulation, exogenous variables influence all future parameters through both the autoregressive term and the score term. In the latter model formulation, they affect future parameters only through the score term. When no exogenous variables are included, the two specifications are equivalent, although with differently parameterized intercept. When numerically optimizing parameter values, the latter model exhibits faster convergence, thanks to the dissociation of $\omega$ from $\varphi_k$. In the context of ARIMA models, the differences between the two model specifications are discussed by \cite{Hyndman2010}. In the context of a GAS model for ranking data, \citet{Holy2025a} advocates the use of a regression model with dynamic errors in an application to ice hockey results.

Other model specifications can be obtained by imposing various restrictions on $\omega$, $\beta_i$, $\alpha_j$, or $\varphi_k$. In addition, it is possible to have different orders $P$ and $Q$ for individual parameters when multiple parameters are time-varying. Furthermore, the set of exogenous variables can also vary for different parameters.

The recursive nature of $f_t$ necessitates the initialization of the first few elements $f_1, \ldots, f_{\max\{P,Q\}}$. A sensible approach is to set them to the long-term value,
\begin{equation}
\label{eq:init}
\bar{f} = \begin{cases}
\left(1 - \sum_{k=1}^Q \varphi_k \right)^{-1} \left(\omega + \sum_{i=1}^M \beta_i \frac{1}{N} \sum_{t=1}^N x_{ti} \right) & \text{in model \eqref{eq:dynJoint}}, \\
\omega + \sum_{i=1}^M \beta_i \frac{1}{N} \sum_{t=1}^N x_{ti} & \text{in model \eqref{eq:dynSep}}.
\end{cases}
\end{equation}

\subsection{Maximum Likelihood Estimation}

GAS models can be straightforwardly estimated by the maximum likelihood method. Let $\theta = (\omega, \beta_1, \ldots, \beta_M, \alpha_1, \ldots, \alpha_P, \varphi_1, \ldots, \varphi_Q, g)'$ denote the vector of all parameters to be estimated. The estimate $\hat{\theta}$ is then obtained by maximizing the full log-likelihood as
\begin{equation}
\hat{\theta} = \arg\max_{\theta} \sum_{t=1}^T \ln p(y_t|f_t,g).
\end{equation}
Alternatively, the conditional log-likelihood can be maximized, which excludes the initial $\max\{P,Q\}$ terms. The maximization of the log-likelihood function can be accomplished using various general-purpose algorithms designed for solving nonlinear optimization problems.

The standard errors of the estimated parameters can be obtained using the standard maximum likelihood asymptotics. Under appropriate regularity conditions, the maximum likelihood estimator $\hat{\theta}$ is consistent and asymptotically normal. Specifically, it satisfies:
\begin{equation}
\sqrt{T} \big( \hat{\theta} - \theta_0 \big) \stackrel{\mathrm{d}}{\to} \mathrm{N} \big( 0, -\mathrm{E}[H(\theta_0)]^{-1} \big), 
\end{equation}
where $\theta_0$ represents the true parameter values and $H(\theta)$ denotes the Hessian of the log-likelihood, defined as
\begin{equation}
H(\theta) = \frac{\partial^2 \ln p(y_t | f_t, g)}{\partial \theta \partial \theta'}.
\end{equation}

In finite samples, the expected value of the Hessian can be approximated by the empirical Hessian of the log-likelihood evaluated at the estimated parameter values $\hat{\theta}$. This empirical Hessian provides an estimate of the curvature of the log-likelihood function and serves as a practical substitute for the true expected value of the Hessian when finite-sample inference is required.

The conditions for the consistency and asymptotic normality of the estimator depend on the specific distributional assumptions and dynamics of the model and need to be verified on a case-by-case basis. Each distribution may have its own specific characteristics and requirements for maximum likelihood estimation. For the general asymptotic theory regarding GAS models and maximum likelihood estimation, see \cite{Blasques2014a}, \cite{Blasques2018}, and \cite{Blasques2022}.

\subsection{Theoretical and Empirical Properties}

The use of the score for updating time-varying parameters is optimal in an information theoretic sense. For an investigation of the optimality properties of GAS models, see \cite{Blasques2015} and \cite{Blasques2021b}.

Generally, the GAS models perform quite well when compared to alternatives, including parameter-driven models. For a comparison of the GAS models to alternative models, see \cite{Koopman2016} and \cite{Blazsek2020}.

\subsection{Notable Models}

The GAS class includes many well-known econometric models, such as the generalized autoregressive conditional heteroskedasticity (GARCH) model of \cite{Bollerslev1986} based on the normal distribution, the autoregressive conditional duration (ACD) model of \cite{Engle1998} based on the exponential distribution, and the count model of \cite{Davis2003} based on the Poisson distribution. 

More recently, a variety of novel score-driven models has been proposed, such as the Beta-t-(E)GARCH model of \cite{Harvey2008}, a multivariate Student's t volatility model of \cite{Creal2011}, a Dirichlet model of \cite{Calvori2013}, the GRAS copula model of \cite{DeLiraSalvatierra2015}, the realized Wishart-GARCH model of \cite{Hansen2016a}, a bimodal Birnbaum--Saunders model of \cite{Fonseca2018}, a Skellam model of \cite{Koopman2018}, a circular model of \cite{Harvey2024}, a Bradley--Terry model of \cite{Gorgi2019}, a bivariate Poisson model of \cite{Koopman2019}, a censoring model of \cite{Harvey2020}, a double Poisson mixture model of \cite{Holy2022b}, a ranking model of \cite{Holy2022f}, a Tobit model of \cite{Harvey2023}, and a zero-inflated negative binomial model of \cite{Blasques2024a}.

For an overview of various GAS models, see \cite{Artemova2022} and \cite{Harvey2022}.

\section{Features of the Package}
\label{sec:pkg}

\subsection{Model Specification and Estimation}

The heart of the \pkg{gasmodel} package is the \code{gas()} function, which serves as a powerful tool for estimating both univariate and multivariate GAS models. This function offers extensive flexibility with its wide range of arguments:

\begin{longcode}
gas(y, x = NULL, distr, param = NULL, scaling = "unit", regress = "joint",
  p = 1L, q = 1L, par_static = NULL, par_link = NULL, par_init = NULL,
  lik_skip = 0L, coef_fix_value = NULL, coef_fix_other = NULL,
  coef_fix_special = NULL, coef_bound_lower = NULL, coef_bound_upper = NULL,
  coef_start = NULL, optim_function = wrapper_optim_nloptr, optim_arguments =
  list(opts = list(algorithm = "NLOPT_LN_NELDERMEAD", xtol_rel = 0, maxeval =
  1e+06)), hessian_function = wrapper_hessian_stats, hessian_arguments = list(),
  print_progress = FALSE)
\end{longcode}

However, at its core, it only requires two essential inputs: a time series \code{y} and a distribution \code{distr}. All other arguments come with default values, ensuring that the function can be readily used even with minimal specifications.

A time series \code{y} can be represented as either a vector of length $T$ or a $T \times 1$ matrix in the case of univariate series. In the multivariate case, it should be a $T \times N$ matrix, where $N$ denotes the dimension of the series.

Additionally, there is an option to include exogenous variables \code{x}. When incorporating a single variable that is common for all time-varying parameters, a numeric vector of length $T$ can be provided. For multiple variables that are common for all time-varying parameters, a $T \times M$ numeric matrix can be used. In cases where there are individual variables for each time-varying parameter, a list of numeric vectors or matrices following the aforementioned formats can be utilized. To control whether the variables are included in the dynamics equation together, as in \eqref{eq:dynJoint}, the argument \code{regress} can be set to \code{"joint"}. Alternatively, if separate equations for dynamics and regression are preferred, as in \eqref{eq:dynSep}, \code{regress} can be set to \code{"sep"}. The default behavior is \code{regress\ =\ "joint"}, which is the model users might expect when incorporating exogenous variables. However, in general, we recommend to use \code{regress\ =\ "sep"} to improve the numerical performance of the optimizer and to provide more straightforward interpretability of the model.

The selection of the distribution in the \code{gas()} function is determined by the \code{distr} argument. Some distributions have multiple parametrizations available, which can be specified using the \code{param} argument. It is important to note that certain parameters may have restrictions imposed on them, and these restrictions should be considered in the model dynamics. However, it may not always be possible to satisfy these restrictions, or it may require additional constraints on the coefficients controlling the dynamics. To handle parameter restrictions, it is generally recommended to use a link function that transforms the parameters into unrestricted real numbers. By default, the logarithmic function is applied to time-varying parameters in the interval $(0, \infty)$, while the logistic function is used for time-varying parameters in the interval $(0, 1)$. The static parameters are unaffected. This behavior can be modified by the \code{par_link} argument, which takes the form of a logical vector. The \code{TRUE} values indicate that the logarithmic/logistic link is applied to the corresponding parameters. The list of available distributions and their parametrizations can be obtained using the \code{distr()} function. Alternatively, Table \ref{tab:distr} provides the relevant information.

The determination of time-varying and static parameters is guided by the \code{par_static} argument, which takes the form of a logical vector. The \code{TRUE} values indicate static parameters. By default, the first parameter of the distribution is considered time-varying, while the remaining parameters are treated as static. The score order $P$ and the autoregressive order $Q$ are selected by the \code{p} and \code{q} arguments respectively. These arguments can take either a single non-negative integer or a vector of non-negative integers when different orders are required for different parameters.

The choice of scaling function for the score is determined by the \code{scaling} argument. The supported scaling options include the unit scaling (\code{scaling = "unit"}), the scaling based on the inverse of the Fisher information matrix (\code{scaling = "fisher_inv"}), and the scaling based on the inverse square root of the Fisher information matrix (\code{scaling = "fisher_inv_sqrt"}). The latter two scalings utilize the Fisher information for the time-varying parameters exclusively. If the preference is to use the full Fisher information matrix, which includes both time-varying and static parameters, the \code{"full_fisher_inv"} or \code{"full_fisher_inv_sqrt"} scaling options can be selected. For the individual Fisher information associated with each parameter, the \code{"diag_fisher_inv"} and \code{"diag_fisher_inv_sqrt"} scaling options are available. It should be noted that when the parametrization is orthogonal (see \code{distr()}), there are no differences among these scaling variants.

The first $\max \{ P, Q \}$ initial values of the time-varying parameters are by default set to their long-term values \eqref{eq:init}. It is also possible to assign specific values to the initial parameters using the \code{par_init} argument. During the maximization of the log-likelihood, the initial values can be included, resulting in the computation of the full likelihood, which is the default option. Alternatively, the initial values can be omitted, leading to the computation of the conditional likelihood by specifying \code{lik_skip = NULL}. To exclude a specified number of first few values from the likelihood calculation, a non-negative integer can be provided to \code{lik_skip}.

Restrictions on estimated coefficients can be enforced using several arguments. The \code{coef_fix_value} argument allows coefficients to be fixed at specific values, using a numeric vector where \code{NA} values indicate coefficients that are not fixed. To set coefficients as linear combinations of other coefficients, the \code{coef_fix_other} argument can be used. It requires a square matrix with multiples of the estimated coefficients, which are added to the fixed coefficients. A coefficient given by row is fixed on coefficient given by column. By this logic, all rows corresponding to the estimated coefficients should contain only \code{NA} values. All columns corresponding to the fixed coefficients should also contain only \code{NA} values. For convenience, common coefficient structures can be specified by name using the \code{coef_fix_special} argument. Examples include \code{panel_structure}, \code{zero_sum_intercept}, and \code{random_walk}. To impose lower and upper bounds on coefficients, the \code{coef_bound_lower} and \code{coef_bound_upper} arguments can be utilized, respectively.

The \code{coef_start} argument allows for the specification of the starting values of coefficients used in the optimization process. If no values are provided, the starting values are automatically selected from a small grid of values. To define the optimization function, the \code{optim_function} argument is used. The function should be formatted according to the required specifications. Two wrapper functions are available for convenience: \code{wrapper_optim_stats()}, which utilizes the \code{optim()} function from the \pkg{stats} package, and \code{wrapper_optim_nloptr()}, which utilizes the \code{nloptr()} function from the \pkg{nloptr} package \citep{Ypma2020}. Additional arguments can be passed to the optimization function as a list using the \code{optim_arguments} argument. Similarly, the Hessian matrix can be computed using the function specified in the \code{hessian_function} argument. Three wrappers are available: \code{wrapper_hessian_stats} for the \code{optimHess()} function from the \pkg{stats} package, \code{wrapper_hessian_pracma} for the \code{hessian()} function from the \pkg{pracma} package \citep{Borchers2023}, and \code{wrapper_hessian_numderiv} for the \code{hessian()} function from the \pkg{numDeriv} package \citep{Gilbert2022}. Additional arguments for the Hessian function can be passed as a list using the \code{hessian_arguments} argument. If desired, a detailed computation report can be continuously printed by setting the \code{print_progress} argument to \code{TRUE}.

The function returns a list of S3 class \code{gas}. This list consists of five components: \code{data}, \code{model}, \code{control}, \code{solution}, and \code{fit}, each of which is also a list. The \code{data} component contains the supplied time series and exogenous variables. The \code{model} component contains the specification of the model structure and size. The \code{control} component contains the settings that control the optimization and Hessian computation. The \code{solution} component contains the raw results of the optimization and Hessian computation. Lastly, and most importantly, the \code{fit} component contains comprehensive estimation results. When an object of the \code{gas} class is printed, it provides a concise summary similar to the \code{summary.lm()} function from the \pkg{stats} package. Various generic functions can be applied to \code{gas} objects, including \code{summary()}, \code{plot()}, \code{coef()}, \code{vcov()}, \code{fitted()}, \code{residuals()}, \code{logLik()}, \code{AIC()}, \code{BIC()}, and \code{confint()}.

\subsection{Forecasting}

Forecasting of GAS models is performed using the \code{gas_forecast()} function. This function offers two forecasting methods. The \code{mean_path} method filters the time-varying parameters based on zero score and then generates the mean of the time series. The \code{simulated_paths} method repeatedly simulates time series, simultaneously filters time-varying parameters, and then estimates mean, standard deviation, and quantiles. See \cite{Blasques2016a} for more details on this method.

To use the \code{gas_forecast()} function, an estimated GAS model is required. Typically, the output of the \code{gas()} function (a \code{gas} object) can be supplied via the \code{gas_object} argument:

\begin{longcode}
gas_forecast(gas_object, method = "mean_path", t_ahead = 1L, x_ahead = NULL,
  rep_ahead = 1000L, quant = c(0.025, 0.975))
\end{longcode}

Alternatively, multiple arguments including the data, model specification, and estimated coefficients can be manually specified:

\begin{longcode}
gas_forecast(method = "mean_path", t_ahead = 1L, x_ahead = NULL, 
  rep_ahead = 1000L, quant = c(0.025, 0.975), y, x = NULL, distr, param = NULL,
  scaling = "unit", regress = "joint", p = 1L, q = 1L, par_static = NULL,
  par_link = NULL, par_init = NULL, coef_est = NULL)
\end{longcode}

The forecasting method is determined by the \code{method} argument. The number of observations to forecast can be specified using the \code{t_ahead} argument. If exogenous variables are utilized, their values must be provided for the forecasted period using the \code{x_ahead} argument. For the \code{simulated_paths} method, the number of simulations can be controlled using the \code{rep_ahead} argument, and the desired quantiles can be specified using the \code{quant} argument.

The function returns a list of S3 class \code{gas_forecast} with three components: \code{data}, \code{model}, \code{forecast}. The \code{data} component contains the supplied time series and exogenous variables. The \code{model} component contains the specification of the model structure and size. The \code{forecast} component contains the mean of the forecasted observations, along with standard deviations and quantiles if the \code{simulated_paths} method is used. Available generic functions are \code{summary()} and \code{plot()}.

\subsection{Simulation}

Basic simulation of GAS models is handled by the \code{gas_simulate()} function.

The \code{gas_simulate()} function requires supplying the coefficients using the \code{coef_est} argument and specifying the model using arguments \code{distr}, \code{param}, \code{scaling}, \code{regress}, \code{p}, \code{q}, \code{par_static}, \code{par_link}, \code{par_init}, and, in the case of multivariate models, the dimension \code{n}:

\begin{longcode}
gas_simulate(t_sim = 1L, x_sim = NULL, distr, param = NULL, scaling = "unit",
  regress = "joint", n = NULL, p = 1L, q = 1L, par_static = NULL,
  par_link = NULL, par_init = NULL, coef_est = NULL)
\end{longcode}

Alternatively, only a \code{gas} object containing a model estimated by the \code{gas()} function can be provided using the \code{gas_object} argument:

\begin{longcode}
gas_simulate(gas_object, t_sim = 1L, x_sim = NULL)
\end{longcode}

The number of observations to simulate can be specified using the \code{t_sim} argument. If exogenous variables are utilized, their values must be provided for the simulation sample using the \code{x_sim} argument.

The function returns a list of S3 class \code{gas_simulate} with three components: \code{data}, \code{model}, \code{simulation}. The \code{data} component contains the exogenous variables, if supplied. The \code{model} component contains the specification of the model structure and size. The \code{simulation} component contains the simulated time series, time-varying parameters, and scores. Available generic functions are \code{summary()} and \code{plot()}.

\subsection{Bootstrapping}

To compute standard deviations and confidence intervals of the estimated coefficients in GAS models, the package provides the \code{gas_bootstrap()} function. This function employs the bootstrapping technique to estimate the uncertainty associated with the coefficients. The \code{parametric} method involves repeatedly simulating time series using the parametric model and re-estimating the coefficients based on the simulated data. The \code{simple_block}, \code{moving_block}, and \code{stationary_block} methods execute the circular block bootstrap with fixed non-overlapping blocks, fixed overlapping blocks, and randomly sized overlapping blocks, respectively.

The \code{gas_bootstrap()} function requires an estimated GAS model as input. The best way is to simply supply a \code{gas} object to the \code{gas_object} argument:

\begin{longcode}
gas_bootstrap(gas_object, method = "parametric", rep_boot = 1000L, 
  block_length = NULL, quant = c(0.025, 0.975), optim_function =
  wrapper_optim_nloptr, optim_arguments = list(opts = list(algorithm =
  "NLOPT_LN_NELDERMEAD", xtol_rel = 0, maxeval = 1e+04)), parallel_function =
  NULL, parallel_arguments = list())
\end{longcode}

Alternatively, the individual arguments including the data, model specification, and estimated coefficients, can be provided:

\begin{longcode}
gas_bootstrap(method = "parametric", rep_boot = 1000L, block_length = NULL,
  quant = c(0.025, 0.975), y, x = NULL, distr, param = NULL, scaling = "unit",
  regress = "joint", p = 1L, q = 1L, par_static = NULL, par_link = NULL,
  par_init = NULL, lik_skip = 0L, coef_fix_value = NULL, coef_fix_other = NULL,
  coef_fix_special = NULL, coef_bound_lower = NULL, coef_bound_upper = NULL,
  coef_est = NULL, optim_function = wrapper_optim_nloptr, optim_arguments = 
  list(opts = list(algorithm = "NLOPT_LN_NELDERMEAD", xtol_rel = 0, maxeval = 
  1e+06)), parallel_function = NULL, parallel_arguments = list())
\end{longcode}

The bootstrapping method is determined by the \code{method} argument. The number of bootstrap samples is specified by the \code{rep_boot} argument. For the \code{simple_block} and \code{moving_block} methods, the fixed size of blocks must be specified by the \code{block_length} argument. For the \code{stationary_block} method, the mean size of blocks must be specified by the \code{block_length} argument. The desired quantiles can be specified using the \code{quant} argument. As bootstrapping can be computationally very demanding, parallelization is achievable by employing the \code{parallel_function} argument, which expects a function similar to \code{lapply()}, allowing the application of a function over a list. Two wrapper functions are available for convenience: \code{wrapper_parallel_multicore()}, which utilizes the multicore parallelization functionality from the \pkg{parallel} package, and \code{wrapper_parallel_snow()}, which utilizes the snow parallelization functionality from the \pkg{parallel} package. Additional arguments can be passed to the parallelization function as a list using the \code{parallel_arguments} argument. If \code{parallel_function} is set to \code{NULL}, no parallelization is employed and \code{lapply()} is used. 

The function returns a list of S3 class \code{gas_bootstrap} with three components: \code{data}, \code{model}, \code{bootstrap}. The \code{data} component contains the supplied time series and exogenous variables. The \code{model} component contains the specification of the model structure and size. The \code{bootstrap} component contains the full set of bootstrapped coefficients as well as the basic statistics derived from them. Available generic functions are \code{summary()}, \code{plot()}, \code{coef()}, and \code{vcov()}.

\subsection{Filtered Parameters}

The filtered time-varying parameters of an estimated model can be directly obtained from the output of the \code{gas()} function. However, to investigate the uncertainty associated with these parameters, the \code{gas_filter()} function can be used. This function also supports forecasting and provides two methods. The \code{simulated_coefs} method calculates a path of time-varying parameters for each simulated coefficient set, assuming asymptotic normality with a given variance-covariance matrix. See \cite{Blasques2016a} for more details on this method. The \code{given_coefs} methods computes a path of time-varying parameters for each supplied coefficient set. Suitable sets of coefficients can be obtained, for example, through the use of the \code{gas_bootstrap()} function.

An estimated GAS model can be supplied as a \code{gas} object to the \code{gas_object} argument:

\begin{longcode}
gas_filter(gas_object, method = "simulated_coefs", coef_set = NULL,
  rep_gen = 1000L, t_ahead = 0L, x_ahead = NULL, rep_ahead = 1000L,
  quant = c(0.025, 0.975))
\end{longcode}

Alternatively, the individual arguments including the data, model specification, and estimated coefficients with variance-covariance matrix can be provided:

\begin{longcode}
gas_filter(method = "simulated_coefs", coef_set = NULL, rep_gen = 1000L,
  t_ahead = 0L, x_ahead = NULL, rep_ahead = 1000L, quant = c(0.025, 0.975), y,
  x = NULL, distr, param = NULL, scaling = "unit", regress = "joint", p = 1L,
  q = 1L, par_static = NULL, par_link = NULL, par_init = NULL,
  coef_fix_value = NULL, coef_fix_other = NULL, coef_fix_special = NULL,
  coef_bound_lower = NULL, coef_bound_upper = NULL, coef_est = NULL,
  coef_vcov = NULL)
\end{longcode}

The \code{method} argument determines the approach for capturing uncertainty. For the \code{given_coefs} method, the \code{coef_set} argument in the form a numeric matrix of coefficient sets in rows must be provided. For the \code{simulated_coefs} method, the \code{rep_gen} argument representing the number of generated coefficient sets must be provided. If forecasting is desired, the number of observations to forecast can be specified using the \code{t_ahead} argument, values of exogenous variable for the forecasted period can be provided using the \code{x_ahead} argument, and the number of simulation repetitions in the forecasted sample can be controlled using the \code{rep_ahead} argument. The desired quantiles can be specified using the \code{quant} argument.

The function returns a list of S3 class \code{gas_filter} with three components: \code{data}, \code{model}, \code{filter}. The \code{data} component contains the supplied time series and exogenous variables. The \code{model} component contains the specification of the model structure and size. The \code{filter} component contains in-sample and possibly out-of-sample means, standard deviations, and quantiles of the time-varying parameters and scores. Available generic functions are \code{summary()} and \code{plot()}.

\subsection{Supplementary Functions for Distributions}

The \code{distr()} function can be utilized to retrieve a list of distributions and their parametrizations supported by the \code{gas()} function. To narrow down the output and focus on specific distributions, arguments such as \code{filter_type} or \code{filter_dim}, among others, can be specified. The output is in the form of a data.frame with columns providing information on the distributions such as the data type, dimension, orthogonality, and default parameterization.

To work with individual distributions, the \pkg{gasmodel} package offeres several functions. The \code{distr_density()} function computes the density of a given distribution, the \code{distr_mean()} function computes the mean of a given distribution, the \code{distr_var()} function computes the variance of a given distribution, the \code{distr_score()} function computes the score of a given distribution, the \code{distr_fisher()} function computes the Fisher information of a given distribution, and the \code{distr_random()} function generates random observations from a given distribution. Each of these function can be supplied with arguments specifying the distribution and the parametrization, namely \code{distr}, \code{param}, \code{par_link}. It is important to note that while the \code{gas()} function may automatically set the logarithmic/logistic link for time-varying parameters, it must be set manually for the distribution functions. Additionaly, a vector of parameter values must be provided to the \code{f} argument. Some functions may also require an observation to be provided to the \code{y} argument. For detailed usage instructions, please refer to the documentation for each individual function.

\section{Example Usage}
\label{sec:case}

\subsection{Analysis of Toilet Paper Sales}

To demonstrate the practical application of the package, we conduct an analysis on the dynamics of toilet paper sales. The dataset \code{toilet_paper_sales} includes the daily number of toilet paper packs sold in a European store during the years 2001 and 2002, along with a promo variable indicating whether the product was featured in a campaign. In addition to the promo dummy variable, we utilize dummy variables for each day of the week, which are already supplied in the dataset.

There are some missing values, corresponding to the days when the store was closed. It is not necessary to remove these values as the \code{gas()} function is capable of handling them. When missing observations occur, the time-varying parameters are reinitialized---parameter values are set to their initial value (either the unconditional value or the value supplied by the \code{par_init} argument), and the score is set to zero. Naturally, missing observations are excluded from the evaluation of the likelihood function.

\begin{longcode}
data("toilet_paper_sales")
y <- toilet_paper_sales$quantity
x <- as.matrix(toilet_paper_sales[3:9])
\end{longcode}

Given that our primary variable of interest is in the form of counts, we employ a count distribution. The \code{distr()} function can be used to obtain a list of appropriate distributions.

\begin{longcode}
distr(filter_type = "count", filter_dim = "uni", filter_default = TRUE)$distr
\end{longcode}

\begin{longcode}
#> [1] "dpois"    "geom"     "negbin"   "pois"     "zigeom"   "zinegbin" "zipois"
\end{longcode}

We start our analysis by utilizing the Poisson distribution, which is the customary initial choice in count data analysis. The promo and day of the week dummy variables are included as exogenous variables. The dynamics are specified as a regression model with dynamic errors. We estimate the model by the \code{gas()} function.

\begin{longcode}
est_pois <- gas(y = y, x = x, distr = "pois", regress = "sep")
est_pois
\end{longcode}

\begin{longcode}
#> GAS Model: Poisson Distribution / Mean Parametrization / Unit Scaling 
#> 
#> Coefficients: 
#>                    Estimate Std. Error   Z-Test Pr(>|Z|)    
#> log(mean)_omega   2.6517736  0.0436953  60.6878  < 2e-16 ***
#> log(mean)_beta1  -0.0015755  0.0264302  -0.0596  0.95247    
#> log(mean)_beta2  -0.0792304  0.0271167  -2.9218  0.00348 ** 
#> log(mean)_beta3  -0.0108654  0.0265633  -0.4090  0.68251    
#> log(mean)_beta4   0.0554222  0.0257702   2.1506  0.03151 *  
#> log(mean)_beta5  -0.8179358  0.0352252 -23.2202  < 2e-16 ***
#> log(mean)_beta6  -1.7635113  0.0538292 -32.7612  < 2e-16 ***
#> log(mean)_beta7   0.7063464  0.0361533  19.5375  < 2e-16 ***
#> log(mean)_alpha1  0.0142384  0.0012708  11.2042  < 2e-16 ***
#> log(mean)_phi1    0.9818186  0.0121071  81.0946  < 2e-16 ***
#> ---
#> Signif. codes:  0 '***' 0.001 '**' 0.01 '*' 0.05 '.' 0.1 ' ' 1
#> 
#> Log-Likelihood: -2111.266, AIC: 4242.532, BIC: 4288.128
\end{longcode}

The output of the \code{gas()} function follows the following convention for naming coefficients. The first part of a coefficient name (before the underscore) refers to the parameter of the distribution and its transformation. The names of the individual distribution parameters are listed in the \code{distributions} vignette. For nonnegative parameters, the logarithmic transformation, \code{log()}, can be applied, while for nonnegative parameters less than 1, the logit transformation, \code{logit()}, can be used. The second part of the coefficient name (after the underscore) refers to either the constant (\code{omega}), the regression coefficient (\code{beta}), the score coefficient (\code{alpha}), or the autoregressive coefficient (\code{phi}), as described in \eqref{eq:dynBasic}, \eqref{eq:dynJoint}, and \eqref{eq:dynSep}. For \code{beta}, the number denotes the corresponding exogenous variable. For \texttt{alpha} and \texttt{phi}, the number denotes the lag.

The Poisson distribution can be quite restrictive in practice as it necessitates equidispersion, meaning that the mean of the variable is equal to the variance. Subsequently, we opt for a more versatile distribution---the negative binomial distribution, which accommodates overdispersion, allowing the variance to be greater than the mean.

\begin{longcode}
est_negbin <- gas(y = y, x = x, distr = "negbin", regress = "sep")
est_negbin
\end{longcode}

\begin{longcode}
#> GAS Model: Negative Binomial Distribution / NB2 Parametrization / Unit Scaling 
#> 
#> Coefficients: 
#>                    Estimate Std. Error   Z-Test  Pr(>|Z|)    
#> log(mean)_omega   2.6384526  0.0562032  46.9449 < 2.2e-16 ***
#> log(mean)_beta1  -0.0100824  0.0357634  -0.2819   0.77800    
#> log(mean)_beta2  -0.0772088  0.0366240  -2.1081   0.03502 *  
#> log(mean)_beta3  -0.0155342  0.0361813  -0.4293   0.66767    
#> log(mean)_beta4   0.0452482  0.0357004   1.2674   0.20500    
#> log(mean)_beta5  -0.8240699  0.0430441 -19.1448 < 2.2e-16 ***
#> log(mean)_beta6  -1.7736613  0.0589493 -30.0879 < 2.2e-16 ***
#> log(mean)_beta7   0.7037864  0.0481655  14.6118 < 2.2e-16 ***
#> log(mean)_alpha1  0.0256734  0.0035329   7.2670 3.676e-13 ***
#> log(mean)_phi1    0.9769718  0.0175857  55.5549 < 2.2e-16 ***
#> dispersion        0.0349699  0.0051488   6.7918 1.107e-11 ***
#> ---
#> Signif. codes:  0 '***' 0.001 '**' 0.01 '*' 0.05 '.' 0.1 ' ' 1
#> 
#> Log-Likelihood: -2059.834, AIC: 4141.668, BIC: 4191.824
\end{longcode}

We compare the models based on the Poisson and negative binomial distributions using the Akaike information criterion (AIC).

\begin{longcode}
AIC(est_pois, est_negbin)
\end{longcode}

\begin{longcode}
#>            df      AIC
#> est_pois   10 4242.532
#> est_negbin 11 4141.668
\end{longcode}

\begin{figure}
\centering
\includegraphics[width=15cm]{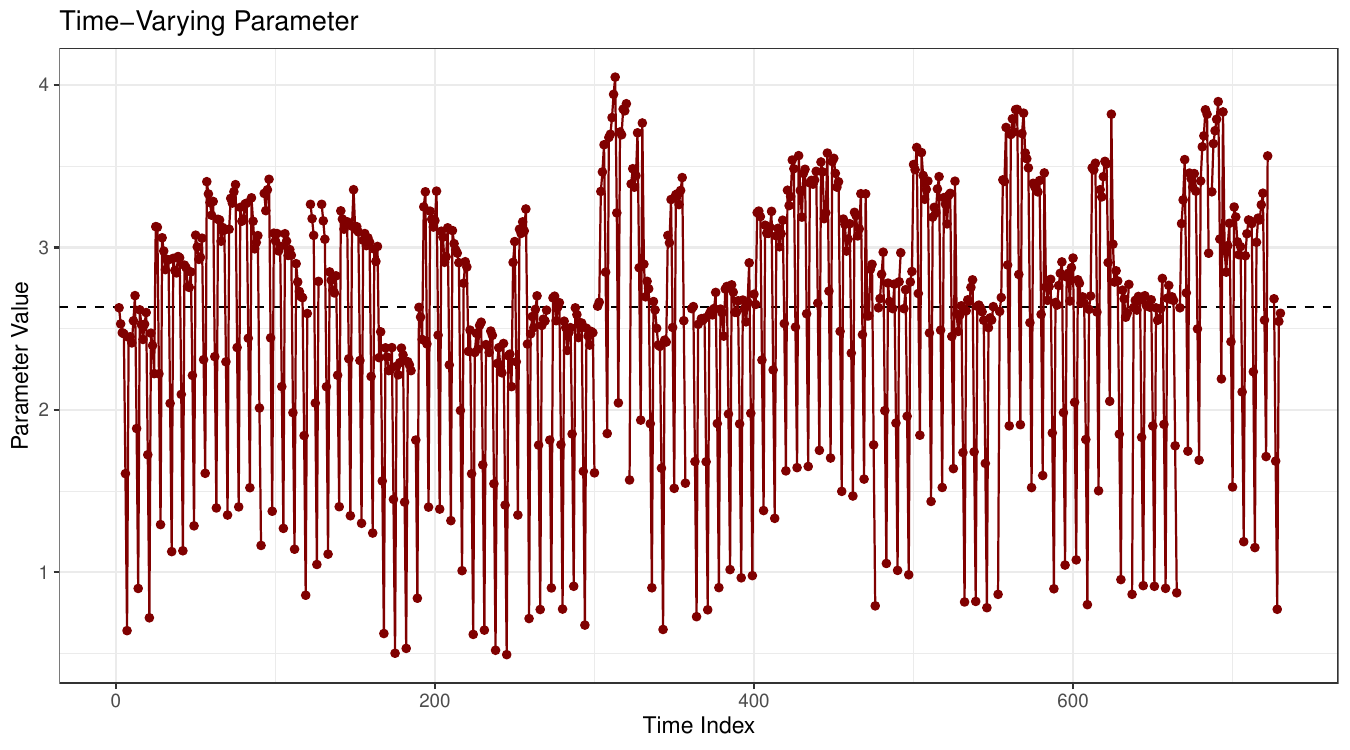}
\caption{The fitted logarithm of the mean. The horizontal dashed line represents the unconditional value of the logarithm of the mean.}
\label{fig:tv}
\end{figure}

Based on the AIC, the negative binomial model is preferred, as the addition of the overdispersion parameter distinctly improves the fit. Nevertheless, the coefficient estimates, standard errors, and p-values are quite similar in both models. The coefficients for working days are close to zero. Note that Monday is treated as the baseline. In contrast, Saturdays and, even more so, Sundays exhibit a significant drop in sales. Promoting the product in a campaign significantly increases sales. The coefficient estimates themselves need to be interpreted with respect to the utilized logarithmic link function. The score and autoregressive coefficients have the expected positive signs. Furthermore, the autoregressive coefficient is close to one, suggesting high persistence. In Figure \ref{fig:tv}, we visualize the time-varying mean with the logarithmic transformation using the \code{plot()} generic function.

\begin{longcode}
plot(est_negbin)
\end{longcode}

\begin{figure}
\centering
\includegraphics[width=15cm]{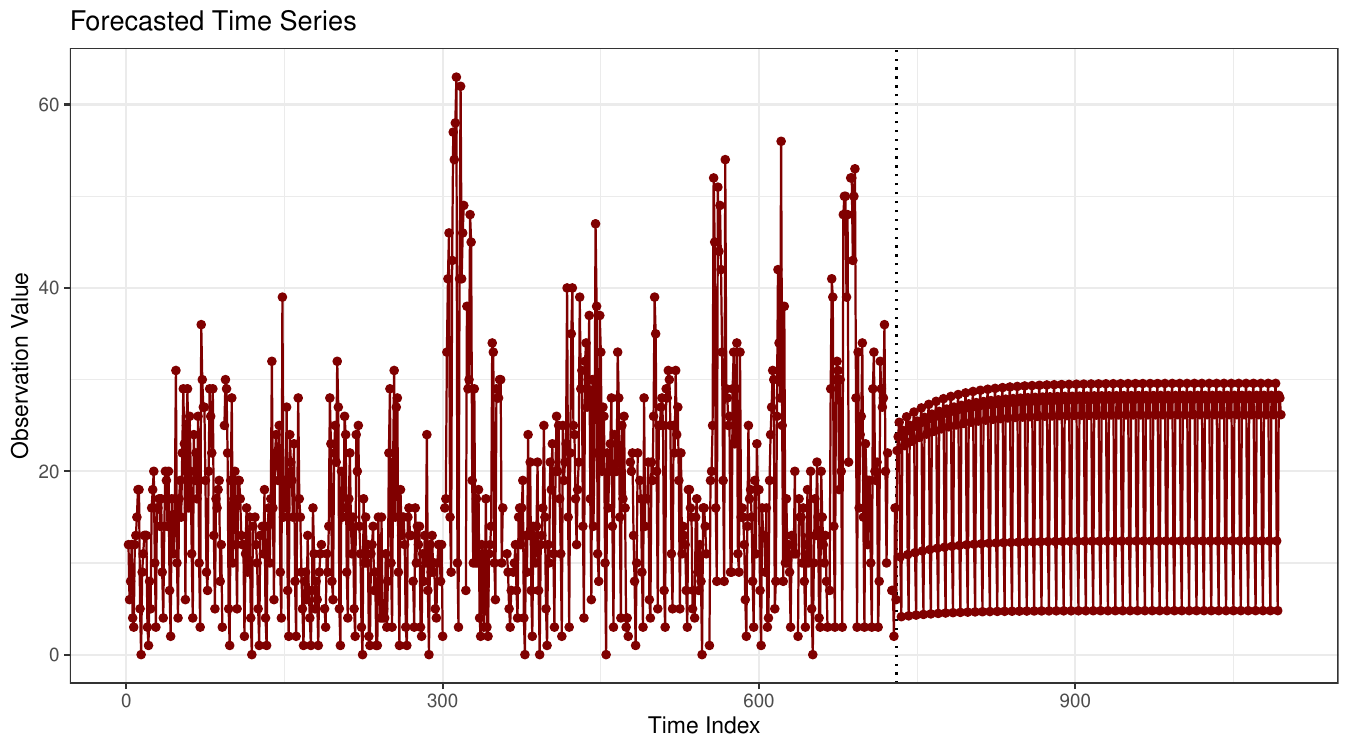}
\caption{The forecasted toilet paper sales for the next year. The product is assumed to be always promoted. The vertical dotted line divides the in-sample and out-of-sample periods.}
\label{fig:fcst}
\end{figure}

Next, we project sales for the following year utilizing the \code{forecast()} function with the default \code{mean_path} method. This method computes time-varying parameters assuming zero score and subsequently generates the mean of the time series. By manipulating the promo dummy variable, we can perform a what-if analysis. In Figure \ref{fig:fcst}, we illustrate the forecasted sales, considering the promotion of the product throughout the entire year.

\begin{longcode}
x_ahead <- cbind(kronecker(matrix(1, 53, 1), diag(7)), 1)[3:367, -1]
fcst_negbin <- gas_forecast(est_negbin, t_ahead = 365, x_ahead = x_ahead)
plot(fcst_negbin)
\end{longcode}

The estimated coefficients are subject to uncertainty. The \code{gas()} function reports standard errors and p-values based on the empirical Hessian justified by the asymptotic theory. However, if the sample size is small, it may be appropriate to derive the standard errors and p-values using the bootstrap method. The \code{gas_bootstrap()} function offers both parametric and block bootstrap options. This process could be computationally intensive, depending on factors such as the number of repetitions, the quantity of observations, the complexity of the model structure, and the optimizer used. To expedite the computation, the function supports parallelization through the arguments \code{parallel_function} and \code{parallel_arguments}. As a simple illustration, we demonstrate the parametric bootstrap with only 10 repetitions. To obtain more accurate estimates of the tails and confidence intervals, it is recommended to use a few thousand repetitions.

\begin{longcode}
set.seed(42)
boot_negbin <- gas_bootstrap(est_negbin, rep_boot = 10)
\end{longcode}

We have a relatively large sample, and the bootstrapped standard errors closely align with those obtained from the empirical Hessian. Note that standard deviations can also be obtained using the \code{vcov()} generic function for both \code{est_negbin} and \code{boot_negbin}.

\begin{longcode}
est_negbin$fit$coef_sd - boot_negbin$bootstrap$coef_sd
\end{longcode}

\begin{longcode}
#>  log(mean)_omega  log(mean)_beta1  log(mean)_beta2  log(mean)_beta3 
#>    -0.0126389857     0.0062104423    -0.0020216634    -0.0040419494 
#>  log(mean)_beta4  log(mean)_beta5  log(mean)_beta6  log(mean)_beta7 
#>     0.0040559045     0.0186693135    -0.0017401544     0.0033140979 
#> log(mean)_alpha1   log(mean)_phi1       dispersion 
#>    -0.0009915867     0.0058394839     0.0008697136
\end{longcode}

As the estimated coefficients are subject to uncertainty, the fitted values of time-varying parameters are also uncertain. To acquire the standard errors and confidence bands of the logarithm of the mean, we employ the \code{gas_filter()} function with the default \code{simulated_coefs} method. This method calculates a path of time-varying parameters for each simulated coefficient set under the assumption of asymptotic normality with a given variance-covariance matrix.

\begin{longcode}
set.seed(42)
flt_negbin <- gas_filter(est_negbin, rep_gen = 100)
\end{longcode}

In our case, the confidence bands are relatively narrow, averaging 0.16 in width.

\begin{longcode}
mean(diff(t(flt_negbin$filter$par_tv_quant[, 1, ])))
\end{longcode}

\begin{longcode}
#> [1] 0.1560328
\end{longcode}

\subsection{Other Case Studies}

For another example of using the package, we refer to the \code{case_durations} vignette, which loosely follows the paper by \cite{Tomanova2021}. This case study examines the timing of bookshop orders in the fashion of autoregressive conditional duration (ACD) models. It employs the generalized gamma distribution with dynamic scale parameter.

The \code{case_ranking} vignette replicates the case study by \cite{Holy2022f}. It analyzes the annual results of the Ice Hockey World Championships, which take the form of time-varying rankings. The analysis utilizes both stationary and random walk specifications, along with the Plackett--Luce distribution incorporating dynamic worth parameters.

\subsection{Limitations and Customization}

There are many GAS models with distributions or structures that are not supported by the \pkg{gasmodel} package. These include copula models \citep{DeLiraSalvatierra2015}, matrix models \citep{Opschoor2018}, censoring models \citep{Harvey2023}, models with parameter interactions \citep{Holy2022b}, sports rating models \citep{Gorgi2019}, semiparametric models \citep{Blasques2016b}, Markov regime-switching models \citep{Bazzi2017}, and spatio-temporal models \citep{Catania2017}. The vignette \code{customization} further discusses these limitations and demonstrates how the package can be customized to accommodate some of these models.

\section{Conclusion}
\label{sec:con}

The purpose of the \pkg{gasmodel} package is to provide researchers, analysts, and data scientists with a versatile toolkit for a broad spectrum of GAS models in R. While it is important to note that not all GAS models found in the literature are supported by the package due to their diverse nature, the package still provides a solid foundation. For some specific GAS models, modifications of the package may be required, or an alternative specialized package/code may prove to be a better option. Nevertheless, the \pkg{gasmodel} package offers considerable flexibility for specifying dynamics, and it boasts an extensive array of probability distribution options.

\section*{Funding}
\label{sec:fund}

The work on this paper was supported by the Czech Science Foundation under project 23-06139S and the personal and professional development support program of the Faculty of Informatics and Statistics, Prague University of Economics and Business.

%\bibliographystyle{myjss}
%\bibliography{RJreferences.bib}

\end{document}